\documentclass[journal=acsodf,manuscript=article]{achemso}

\usepackage{amsmath}
\usepackage{color}
\usepackage{indentfirst}

\author{Yasuhiro Oishi}
\affiliation[Osaka University]
{Graduate School of Engineering Science, Osaka University, 1-3 Machikaneyama-cho, Toyonaka, Osaka 560-8531, Japan}
\email{ooishi.y@opt.mp.es.osaka-u.ac.jp}

\author{Hirotsugu Ogi}
\affiliation[Osaka University]
{Graduate School of Engineering, Osaka University, Suita, Osaka 565-0871, Japan}

\author{Satoshi Hagiwara}
\affiliation[University of Tsukuba]
{Center for Computational Sciences, University of Tsukuba, 1-1-1, Tenno-dai, Tsukuba, Ibaraki 305-8577, Japan}

\author{Minoru Otani}
\affiliation[University of Tsukuba]
{Center for Computational Sciences, University of Tsukuba, 1-1-1, Tenno-dai, Tsukuba, Ibaraki 305-8577, Japan}

\author{Koichi Kusakabe}
\affiliation[University of Hyogo]
{Graduate School of Science, University of Hyogo, 3-2-1 Koto, Kamigori-cho, Ako, Hyogo 678-1297, Japan}

\title[An \textsf{achemso} demo]
  {Theoretical Analysis on the Stability of 1-Pyrenebutanoic Acid Succinimidyl Ester Adsorbed on Graphene}


\begin{document}

\begin{tocentry}

\includegraphics[bb=0 0 600 107, scale = 1]{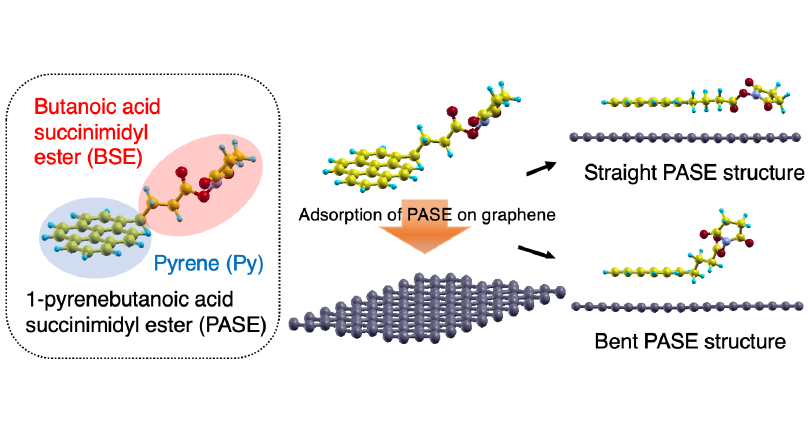}
\label{For Table of Contents Only}

\end{tocentry}
\begin{abstract}
The adsorbed structure of 1-pyrenebutanoic acid succinimidyl ester (PASE) on graphene was investigated based on density functional theory. We found two locally stable structures: a straight structure with the chainlike part of butanoic acid succinimidyl ester (BSE) lying down and a bent structure with the BSE part directed away from graphene, keeping the pyrene (Py) part adsorbed on graphene. Then, to elucidate the adsorption mechanism, we separately estimated the contributions of the Py and BSE parts to the entire PASE adsorption, and the adsorption effect of the BSE part was found to be secondary in comparison to the contribution of the Py. Next, the mobility of the BSE part at room temperature was confirmed by the activation energy barrier between straight and bent structures. To take account of the external environment, we considered the presence of amino acids and the hydration effect by a three-dimensional reference interaction site model. The contributions of glycine molecules and the solvent environment to stabilizing the bent PASE structure relative to the straight PASE structure were found. Therefore, the effect of the external environment around PASE is of importance when the standing-up process of the BSE part from graphene is considered.

\end{abstract}

\section{Introduction}

Graphene, a single-atom-thick and two-dimensional carbon material, has attracted considerable application attention for biosensing devices\cite{mohanty2008graphene,neto2009electronic,georgakilas2012functionalization,wang2011graphene,park2011enhanced,myung2011graphene,yang2010carbon}. Due to its excellent characteristics, such as wide detection area, high carrier mobility, and large heat conductance\cite{cite-key,novoselov2004electric}, graphene is an ideal candidate for biosensor substrates\cite{yang2010carbon}.

One of the important steps in the application of graphene for biosensing devices is the functionalization of graphene with receptors such as antibodies.
Covalent functionalization of graphene is a common approach\cite{loh2010chemistry}.
However, this approach is known to disturb the electronic properties of graphene\cite{georgakilas2012functionalization}.
Meanwhile, noncovalent functionalization, which utilizes $\pi$--$\pi$ stacking, is often employed, as the electronic properties of graphene can be preserved\cite{chen2001noncovalent,kodali2011nonperturbative,liu2012strategies}. 
In the noncovalent approach, linker molecules containing the aromatic group are often used to connect graphene and the receptor protein.

A commonly used linker molecule is 1-pyrenebutanoic acid succinimidyl ester (PASE), which comprises the pyrene (Py) and chainlike parts made of butanoic acid succinimidyl ester (BSE), as shown in Figure\ 1. 
\begin{figure}[t]
    \centering
     \includegraphics[bb=-40 0 600 170, scale = 1]{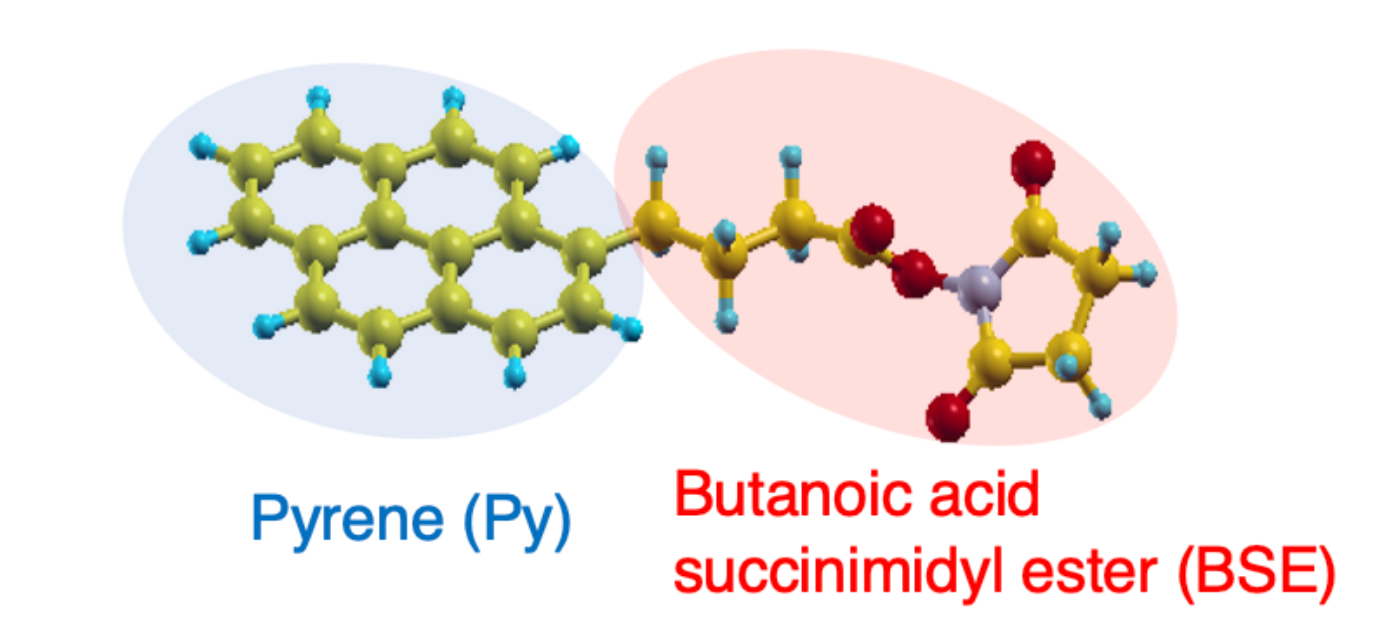}
    \caption{Structure of 1-pyrenebutanoic acid succinimidyl ester (PASE). Atomic geometries are visualized by Xcrysden\cite{kokalj1999xcrysden} in this study (yellow, C; blue, H; red, O; gray, N).}
\label{fig:pase}
\end{figure}
The succinimide group in BSE part reacts with amines in a protein by forming an amide bond, whereas the Py part is considered to noncovalently bind to graphitic materials via $\pi$--$\pi$ stacking\cite{katz1994application,jaegfeldt1983electrochemical,zhou2017label,7388878}. 
PASE linkers are used in phonon biosensors together with the graphene substrate. 
The direct determination of phonon velocities in substrate materials affected by adsorbates is sufficiently sensitive to detect adsorbed biomaterials. 
Recently, based on this mechanism, a biosensing graphene device has been fabricated using a picosecond ultrasonic spectroscopy method\cite{graphenebiosensor2021patent,graphenebiosensor2021proceedings}. 

In measurement analysis, the detailed adsorption mechanism for the linker molecule is desired. Because the vibrational properties of this system are significantly affected by microscopic material structures such as conformations of the linker, understanding the adsorption mechanism of PASE on graphene is necessary to improve the biosensing device.

The Py adsorption mechanism on graphene determines the adsorption of PASE on graphitic materials.
According to a previous study\cite{singh2015noncovalently}, the Py adsorption mainly comes from the van der Waals interaction stabilizing the $\pi$--$\pi$ stacking. On the other hand, the BSE part has been often considered to be always at the site where the amide bond with the receptor protein is formed. Therefore, a standing conformation, {\it i.e.}, the conformation of PASE where the BSE part is standing up from graphene, has been naturally assumed\cite{chen2001noncovalent,katz1994application}.
However, the standing conformation might not be always the case. In other words, the standing conformation might appear only when the graphene-PASE adsorption and PASE-protein bindings are formed simultaneously. 

Theoretically, the adsorption of another linker---1-pyrenebutyric acid (PBA)---on graphene has been explored\cite{thodkar2021self,hinnemo2017monolayer,bailey2014study,li2013solvent}.
A larger binding energy was found when the butyric acid side group was bonded to the graphene sheet ($-1.54$ eV) in comparison to another case when the butyric acid chain was directed away from the graphene sheet ($-1.30$ eV)\cite{bailey2014study}.
Similarly, the adsorption of the PASE linker might come from two contributions of adsorption energies: {\it i.e.}, from Py and BSE. Therefore, we need to consider the straight PASE structure, in which both the Py and BSE parts are lying flat on graphene.
Although there have been several theoretical studies on the PASE adsorption on carbon materials, {\it e.g.}, graphene and a single-walled carbon nanotube\cite{bagherzadeh2021real,zakaria2015nanovector,karachevtsev2011noncovalent,fan2008structural}, it is necessary to further understand the mechanism of the PASE adsorption.

In this study, we investigate the stability of the graphene/PASE system based on density functional theory (DFT). We found two locally stable structures: conformation 1, where the BSE part is lying down, and conformation 2, where the BSE part is standing up. We also considered the presence of amino acids around the PASE linker by including glycine molecules in our DFT simulations. To consider the hydration effect on the graphene/PASE system stability, we used a three-dimensional reference interaction site model (3D-RISM)\cite{kovalenko1998three,sato2000self}.

\section{Results and Discussion}

In this section, we present the results of the adsorption mechanism of the PASE linker.
First, we discuss the PASE adsorption structure and adsorption energies obtained by the DFT calculations.
Next, the mobility of the BSE part is investigated by determining the activation barrier in the standing-up process of the BSE from the minimum energy pathway. Then, we discuss the effects of the external environment on graphene/PASE stability by considering the presence of amino acids and the hydration effect by 3D-RISM. Finally, we provide a discussion on the possible method for improving the sensitivity of the phonon biosensor.

\subsection{Adsorption Structure and Energy}

\begin{figure}[t]
    \centering
    \includegraphics[bb=0 0 700 300, scale = 1]{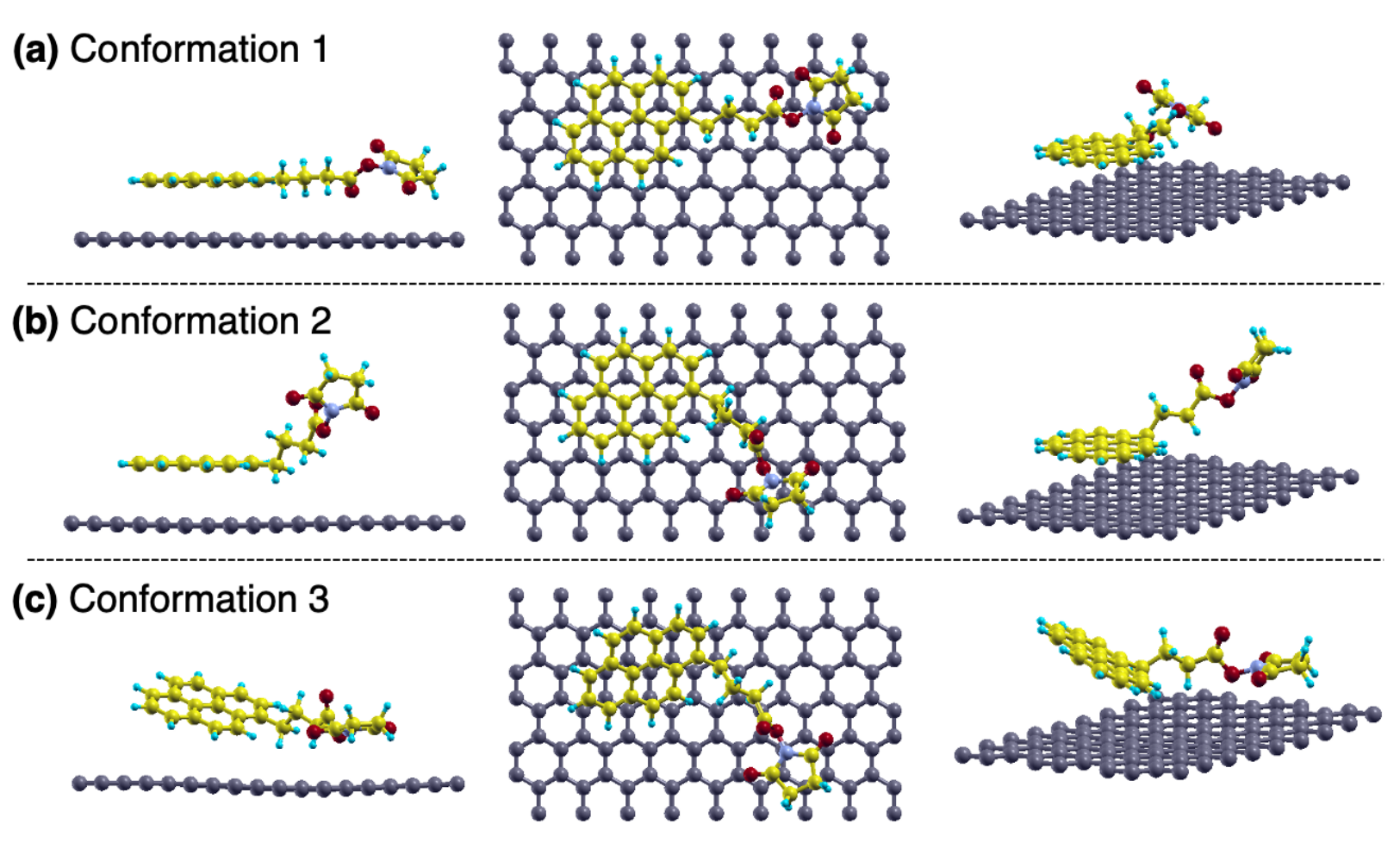}
    \caption{(a--c) Optimized PASE structures on graphene for conformations 1--3, respectively.}
    \label{fig:pase-conf}
\end{figure} 

\indent In Figure\ 2a--c, we show the optimized structures of the PASE on graphene with three conformations. Note that, for conformation 3, the tilted structure of the Py group was fixed during structure optimization. In conformations 1 and 2, the adsorption structures of the Py part on graphene show an AB-stacking, which is consistent with a previous study on Py adsorption on graphene\cite{bailey2014study}. The adsorption energy $E_{\rm ad}(\rm conf,\it i)$ of conformation {\it i} ($i = 1$--$3$) was calculated as
\begin{equation}
E_{\rm ad}(\rm conf,\it i) =  E_{\rm graphene/PASE} - E_{\rm graphene} - E_{\rm PASE}
\end{equation}where $E_{\rm PASE}$ and $E_{\rm graphene}$ denote the total energies of isolated PASE and a graphene sheet, respectively, and $E_{\rm graphene/PASE}$ is the total energy of the PASE adsorbed on graphene. In this definition, a negative value of $E_{\rm ad}(\rm conf,\it i)$ means that the adsorption of PASE is exothermic.

\begin{table}[t]
\caption{Adsorption Energy of Each PASE Conformation $E_{\rm ad}(\rm conf,\it i)$ ($i = 1$--$3$) and Pyrene $E_{\rm ad}(\rm Py)$}
\begin{tabular}{|l|c|} \hline
adsorbate & adsorption energy (eV)  \\ \hline \hline
PASE conformation 1 & $-1.63$   \\
PASE conformation 2 & $-1.28$   \\ 
PASE conformation 3 & $-1.29$  \\ 
pyrene & $-0.88$   \\   \hline                                    
 \end{tabular}
\end{table}

The adsorption energies of each PASE conformation are summarized in Table 1. The absolute value of $E_{\rm ad}(\rm conf,\it i)$ for conformation 1 was larger than that for conformation 2, which is in agreement with a previous study on PBA\cite{bailey2014study}. Thus, the adsorption of PASE on graphene comes from adsorption energies of both Py and BSE.

To further understand the PASE adsorption, we separately estimate the adsorption energies of the Py and BSE parts. For this purpose, an additional calculation on the adsorption energy of only the Py part $E_{\rm ad}(\rm Py)$ was performed, and the result of $E_{\rm ad}(\rm Py)$ is also shown in Table 1. We assume that the total adsorption energy is given by the sum of partial contributions of the Py and BSE parts. Since the total energy difference of isolated PASE between the bent and straight structures was only about 0.003 eV, we neglected the deformation energy in the bent PASE. Thus, we can approximately determine the partial contributions from the Py and BSE parts using the relations

\begin{equation}
E_{\rm ad}(\rm conf,1) = {\it E}_{\rm ad}(\rm Py) + {\it E}_{\rm ad}(\rm BSE)
\end{equation}
\begin{equation}
E_{\rm ad}(\rm conf,2) = {\it E}_{\rm ad}(\rm Py) + {\it E}_{\rm ad}(\rm tilt \, BSE)
\end{equation}
\begin{equation}
E_{\rm ad}(\rm conf,3) = {\it E}_{\rm ad}(\rm tilt \, Py) + {\it E}_{\rm ad}(\rm BSE)
\end{equation}where $E_{\rm ad}(\rm BSE)$, $E_{\rm ad}(\rm tilt \, BSE)$, and $E_{\rm ad}(\rm tilt \, Py)$ denote the adsorption energies of the BSE, tilted BSE, and tilted Py parts, respectively.

From eq\ 2, $E_{\rm ad}(\rm BSE)$ is determined to be $-0.75$ eV, and we also obtain the partial contribution by the tilted parts from eqs\ 3 and 4: $E_{\rm ad}(\rm tilt \, Py) = -0.54$ eV and $E_{\rm ad}(\rm tilt \, BSE) = -0.40$ eV. The absolute value of ${\it E}_{\rm ad}(\rm Py)$ ($E_{\rm ad}(\rm tilt \, Py)$) is larger than that of ${\it E}_{\rm ad}(\rm BSE)$ ($E_{\rm ad}(\rm tilt \, BSE)$), which denotes that the Py mainly contributes to the PASE adsorption. Therefore, the BSE part can easily move while the Py part remains adsorbed on graphene.

\subsection{Mobility of the BSE in the PASE Linker}

\begin{figure}[ht]
    \centering
     \includegraphics[bb=0 0 700 320, scale = 1]{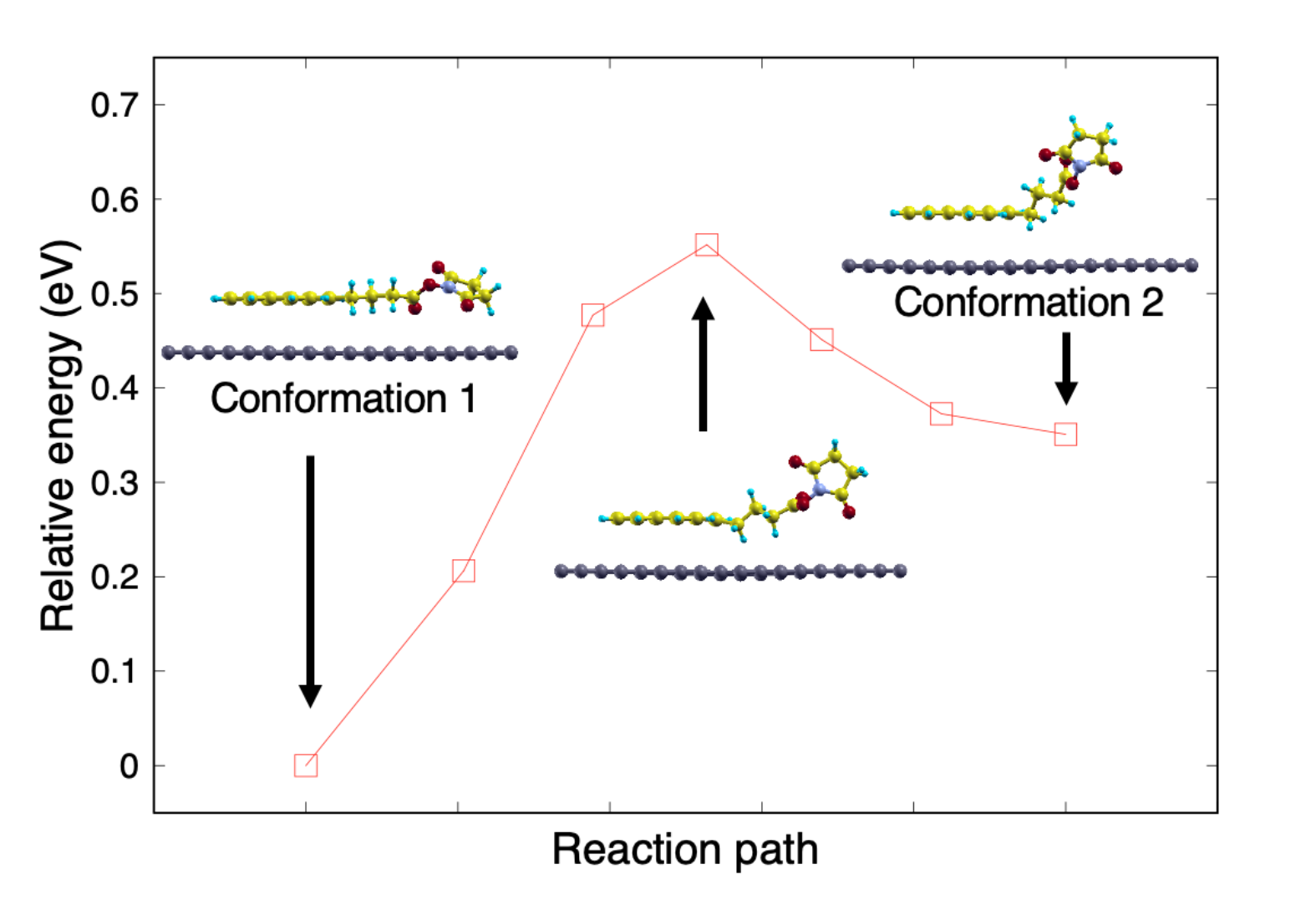}
    \caption{Relative energy of graphene/PASE to the most stable structure (conformation 1) using the nudged elastic band method. The first and final images correspond to optimized structures of conformations 1 and 2, respectively.}
    \label{fig:neb}
\end{figure}

The high conformational flexibility of PASE, {\it i.e.}, the mobility of the BSE part, is thought to be essential to catch a protein approaching from various directions. We discuss the mobility of the BSE part by the activation energy between conformations 1 and 2. To determine the activation energy, we calculated the minimum energy pathway corresponding to the standing-up process of the BSE part by the nudged elastic band method\cite{henkelman2000improved}. The optimized structures of conformations 1 and 2 were used as the initial and final images, respectively, and five intermediate configurations were interpolated.

Figure 3 shows the result of the minimum energy pathway, and the activation energy is 0.20 eV. This result indicates that the transition between conformations 1 and 2 is possible at around room temperature. Therefore, the BSE part can change direction while the Py part remains adsorbed on graphene.

\subsection{Effects of the External Environment}

\begin{figure}[b]
    \centering
     \includegraphics[bb=-7 0 600 150, scale = 1]{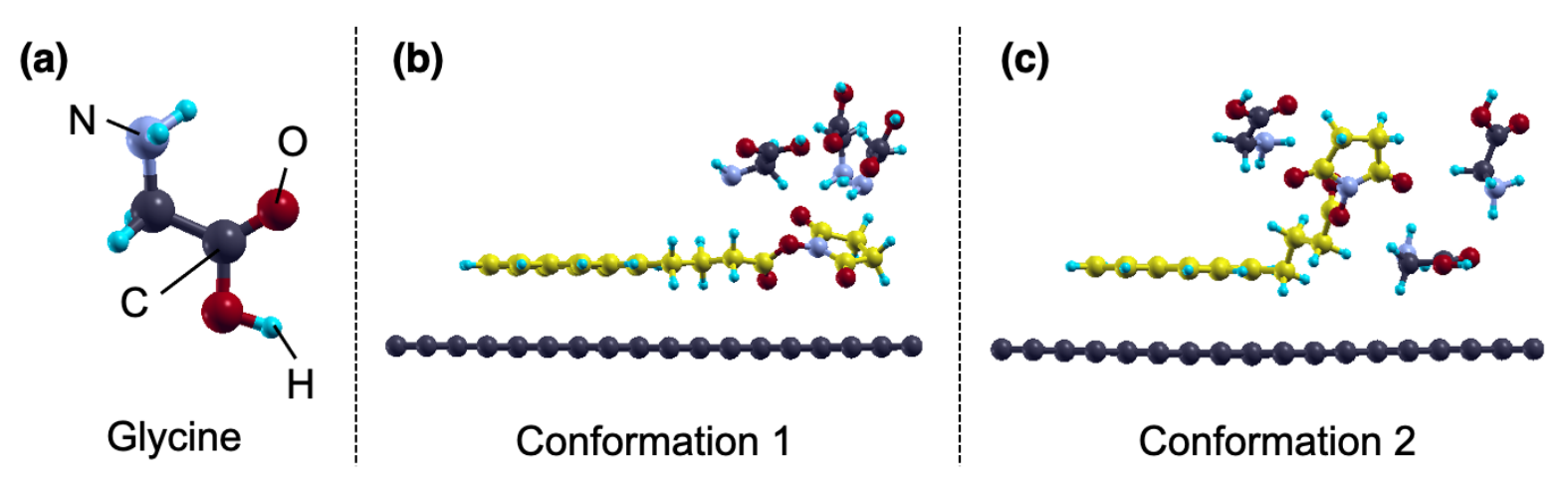}
    \caption{(a) Structure of a glycine molecule. (b, c) Optimized PASE/glycine structures on graphene for conformation 1 and 2, respectively.}
\label{fig:glycine}
\end{figure}

The details of the PASE structure might depend on the operating conditions of the biosensor. When the PASE linker is used for biosensing, proteins coexist with PASE and graphene substrates. In addition, the PASE linker is introduced on graphene substrate by drop-casting. Thus, the PASE linker is often used in aqueous solutions. The external environments emerging under those operating conditions of the biosensor might affect the stability of conformation 2 relative to conformation 1. Therefore, we included these external environment effects in the graphene/PASE system and calculated the total energy of conformation 2 relative to that of conformation 1 (denoted by $\Delta E$).

To reproduce the presence of a protein, we introduced glycine molecules, as shown in Figure\ 4. To take account of the hydration effect, we used 3D-RISM with the Kovalenko and Hirata (KH) closure\cite{kovalenko1999self}, which is a kind of hybrid solvation method combining DFT with classical solution theory (RISM)\cite{hirata1981extended,hirata1982application}. In 3D-RISM, thermodynamic properties such as the solvation free energy can be obtained with lower computational cost in comparison to a molecular dynamics (MD) simulation, which needs many sampling trajectories. The 3D-RISM with the KH closure is known to overestimate the absolute value of the solvation free energy in comparison to the MD simulation and experiment\cite{truchon2014cavity}. However, 3D-RISM describes the hydration energy differences well, and thus we focused on the difference in the hydration free energies between the conformations. In our 3D-RISM simulation, graphene and PASE, as well as glycine molecules, were explicitly treated by DFT, whereas the liquid water was treated as the implicit solution by RISM theory. 

\begin{table}[t]
\caption{Total Energy of Conformation 2 Relative to That of Conformation 1 ($\Delta E$) Obtained by Each Calculation.}
\begin{tabular}{|l|l|l|} \hline
calcd structure & \multicolumn{2}{|l|}{$\Delta E$ (eV)}  \\ \cline{2-3}
                                  & DFT & 3D-RISM  \\ \hline \hline
graphene/PASE  & 0.35 & 0.28 \\ \hline
graphene/PASE/glycine  & 0.09 & $<$ 0.01   \\  \hline                                    
\end{tabular}
\end{table} 

The results of $\Delta E$ are summarized in Table 2. The DFT calculation shows that the glycine molecules increase the stability of conformation 2 relative to conformation 1; $\Delta E$ for graphene/PASE/glycine was only 0.09 eV ($<$ 0.35 eV). The stabilization of conformation 2 may come from the glycine molecule existing between the BSE part and graphene. This glycine prevents the BSE part from lying down on graphene, which can contribute to the stabilization of the standing conformation.

Next, we discuss the results of 3D-RISM and compare them to the results of DFT. For the graphene/PASE system, $\Delta E$ by 3D-RISM was lower than that by DFT, which is also seen in the graphene/PASE/glycine system. The decrease in $\Delta E$ indicates the stabilization of conformation 2 relative to conformation 1 and comes from the hydration effect on PASE adsorbed on graphene. These results suggest that the standing conformation appears more frequently in a solution in comparison to that under vacuum conditions. Thus, the external environment around the PASE linker is of great importance for considering the standing-up process of the BSE part.

\subsection{Possible Methods to Improve a Biosensing Device}
In the phonon biosensors, biomaterials are detected by monitoring the change in the vibrational frequency of the entire resonator on immobilization on the graphene substrate. The vibrational frequency and signal intensity are thought to be significantly affected by the rigidity of the entire adsorbed material, including the linker. The rigidity of the PASE linker will be affected by the detail of the PASE adsorbed structure.
Nevertheless, to the best of our knowledge, there have been no experimental studies that have observed the PASE structure on graphene at an atomic scale. Now that the adsorption mechanism of PASE on graphene has been theoretically elucidated, it will be possible to redesign the rigidity of the linker in order to further increase the signal intensity.

In our simulation, conformation 1 exhibits strong adsorption on graphene. One of the possible ways to increase the rigidity of the PASE linker is to make the PASE a straight form (conformation 1) on graphene while forming an amide bond with the receptor protein. The straight form is expected to be easily realized when the PASE linker is used under vacuum conditions. In the previous section, it was suggested that conformation 2 appears more frequently in a solution in comparison to that under vacuum conditions. Therefore, the signal intensity can be increased by performing detection under vacuum conditions while the PASE and protein are kept immobilized on the graphene substrate. Thus, we can improve the phonon biosensing device.

For other linkers, such as PBA, performing the same analysis as in this study is possible. This kind of study allows us to select the optimal linker that would enhance the sensitivity of the phonon biosensor, which remains as future work.

When the target biomaterial, {\it e.g.}, an antigen, is captured, an amide bond between the PASE linker and receptor protein has to be kept. The strength of the amide bond between the linker and receptor protein has an effect on the vibrational frequency and signal intensity. Therefore, a theoretical evaluation of the bond strength between PASE and the amino acid molecule is necessary. As a typical example, considering the state after the realization of PASE-glycine binding is future work. In order to further elucidate the performance of the PASE linker under the operating conditions of the biosensing device, an observation of the PASE structure using a scanning tunneling microscope or atomic force microscope and biosensor experiments are essential.

\section{Conclusions}

In this study, the adsorbed structure of PASE on graphene was investigated with DFT calculations. Two locally stable structures---a straight PASE with both the Py and chainlike BSE parts lying down on graphene (conformation 1) and a bent PASE with the Py part adsorbed on and the BSE part standing up from graphene (conformation 2)---were found. The absolute value of the adsorption energy of conformation 1 was found to be larger than that of conformation 2 under vacuum conditions. The Py part was found to mainly contribute to the PASE adsorption in comparison to the BSE part. The calculation of the activation energy barrier between conformations 1 and 2 showed that the BSE part can change the direction at room temperature.

We also considered the effects of the external environment around the PASE linker: the presence of amino acids and the hydration effect by 3D-RISM. These external environments were found to contribute to the stabilization of conformation 2 relative to conformation 1. A possible way to improve the phonon biosensor through the redesign of the rigidity of the PASE linker was also discussed. To select an optimal linker that would improve the phonon biosensing device, a theoretical analysis as shown in this paper should be applied to other linkers. The calculation of the PASE-glycine binding, as well as experimental studies such as a scanning tunneling microscope, an atomic force microscope, and biosensor experiments, should be conducted to further clarify the behavior of PASE and its performance in the biosensors.

\section{Computational Method}

The geometrical relaxation was performed on the PASE structure on monolayer graphene using Quantum ESPRESSO\cite{giannozzi2009quantum,giannozzi2017advanced,giannozzi2020quantum}. 
A vacuum region of 10 {\AA} between the adsorbed PASE and the periodic image of the graphene monolayer was found to be large enough to neglect a spurious interaction between adjacent slabs. Therefore, a vacuum region of at least 10 {\AA} in the $z$ direction was inserted in all calculated systems. In the calculation of the graphene/PASE/glycine system, three glycine molecules were placed in the DFT unit cell. An ultrasoft pseudopotential\cite{vanderbilt1990soft} was used for each atom to describe electron-ion interactions. A revised Perdew-Burke-Ernzerhof (PBE) functional for a densely packed solid surface (PBEsol)\cite{perdew2008restoring}, within the generalized gradient approximation (GGA), was used. The cutoff was set to 35 and 350 Ry for wave functions and charge density respectively, and a 2×4×1 Monkhorst-Pack\cite{monkhorst1976special} $\boldsymbol{k}$-point was chosen for all calculations. The convergence threshold on forces was $10^{-6}$ Ry/bohr.

In the 3D-RISM calculation, the periodic boundary condition is imposed on 3D-RISM as well as on the plane-wave and pseudopotential methods\cite{nishihara2017hybrid}. We used liquid water as the solvent around the PASE, and the temperature and solvent density were set to 300 K and 1 g/cm$^3$, respectively. The cutoff energy of the correlation functions was set to 140 Ry, and the KH closure was used. We used a modified simple point charge model\cite{Berendsen1981} for the implicit water, and the universal force field\cite{rappe1992uff} was employed as the Lennard-Jones parameters for the explicit particles.

The semilocal functionals cannot describe the dispersion effect. To take account of the van der Waals interaction, we adopted the DFT-D approach\cite{grimme2006semiempirical,grimme2010consistent,grimme2011effect}.
It has been discussed in the literature\cite{LEBEDEVA201745,PhysRevB.98.174103} that the D3 term used in PBE-D3 and others is effective in assuring accuracy. We validated the use of the DFT-D3 correlation with the PBEsol functional (PBEsol+D3) by determining the equilibrium interplanar distance $d$ between benzene and graphene. In the calculation of the benzene/graphene system, $d$ is determined by calculating the total energy for each interplanar distance. As a reference, we also used the adiabatic-connection fluctuation-dissipation-theorem with random phase approximation (ACFDT-RPA)\cite{PhysRevLett.105.196401} and its extension including the Hubbard correlation energy (ACFDT-RPA+U)\cite{kusakabe2020interplanar}, which is thought to accurately reproduce the van der Waals interaction. The $d$ value between benzene and graphene determined by PBEsol was 3.65 {\AA}, whereas ACFDT-RPA and ACFDT-RPA+U gave 3.46 and 3.44 {\AA}, respectively. PBEsol+D3 gives a $d$ value of 3.40 {\AA}, which is consistent with the ACFDT-RPA(+U) level.

In this study, the structure of the graphene/PASE system was investigated by PBEsol+D3. We also confirmed that a dispersion correction is necessary for structural optimization of the graphene/PASE system. The value of $d$ between the Py and graphene by PBEsol ranges from 3.5 to 3.9 {\AA}. By including the D3 term, the $d$ value between the Py and graphene was shortened, where the value ranges from 3.3 to 3.4 {\AA}. This shortened value is in agreement with the $d$ value for the graphene/benzene system given by PBEsol+D3 (3.40 \AA). For the adsorption energy, PBEsol+D3 gave 1.63 eV for conformation 1 and 1.28 eV for conformation 2. On the other hand, PBEsol showed values deviating from the adsorption energies given by PBEsol+D3; the adsorption energies were 0.41 eV for conformation 1 and 0.36 eV for conformation 2, respectively.


\section*{Author information}

\textbf{Yasuhiro Oishi} - Graduate School of Engineering Science, Osaka University, Toyonaka, Osaka 560-8531, Japan; E-mail: ooishi.y@opt.mp.es.osaka-u.ac.jp

\textbf{Hirotsugu Ogi} - Graduate School of Engineering, Osaka University, Suita, Osaka 565-0871, Japan;
E-mail: ogi@prec.eng.osaka-u.ac.jp

\textbf{Satoshi Hagiwara} - Center for Computational Sciences, University of Tsukuba, Tsukuba, Ibaraki 305-8577, Japan; E-mail: hagiwara@ccs.tsukuba.ac.jp

\textbf{Minoru Otani} - Center for Computational Sciences, University of Tsukuba, Tsukuba, Ibaraki 305-8577, Japan; E-mail: otani@ccs.tsukuba.ac.jp

\textbf{Koichi Kusakabe} - Graduate School of Science, University of Hyogo, Kamigori, Hyogo 678-1297, Japan; E-mail: kusakabe@sci.u-hyogo.ac.jp

\section*{Notes}
The authors declare no competing financial interest.

\begin{acknowledgement}
Y. O. and K. K. thank Y. Wicaksono, N. Morishita, S. Akiyama, and R. Ouchi for illuminating discussions and valuable comments. All calculations were done at the computer centers of Kyushu University and ISSP, University of Tokyo. This work was partially supported by JSPS KAKENHI Grant Numbers JP19H00862 and JP22K04864.
\end{acknowledgement}

\bibliography{pase-gra.bib}

\end{document}